\documentclass[twoside]{article}
\usepackage{din_a4,capt2,PHYSMAC}
\usepackage{subfigure} \usepackage{epsfig}

\begin{document}
{\pagestyle{empty}
\setlength{\oddsidemargin}{0pt}
\setlength{\evensidemargin}{0pt}
\setlength{\marginparwidth}{2.4truecm}
\setlength{\marginparsep}{0pt}
\setlength{\topmargin}{0pt}
\setlength{\headheight}{0pt}
\setlength{\headsep}{0pt}
\setlength{\topskip}{0pt}
\setlength{\textwidth}{16.4truecm}
\renewcommand{\baselinestretch}{1.0}
\parskip = 6pt plus 0.10fill
\let\thepage\relax
\def\centerformat{%
    \pretolerance = 2000 \tolerance = 3000 \hbadness 500
    \parindent = 0pt \parskip = 0pt \catcode"0D = 5
    \rightskip = 20pt plus 1fill\leftskip  = 20pt plus 1fill
    \spaceskip = 0.35em \xspaceskip = 0.45em \parfillskip = 0pt
    \hyphenpenalty = 1000 \exhyphenpenalty = 1000}
\def\:{.\kern 2pt\relax}
\font\sixrm=cmr6
\font\egtrm=cmr8
\def\title#1{{\centerformat{%
    \Large \bf
    #1\par}}%
    \vskip 2pt\vskip\parskip}
\def\authors#1{{\centerformat{\rm #1\par}}}
\def\inst#1{{\centerformat{\it #1\par}}\vskip \parskip}
\newcount\fnnumber
\def\fn#1#2{\/%
    \expandafter\ifx\csname FnN#1\endcsname\relax%
    \global\advance\fnnumber 1%
    \expandafter\xdef\csname FnN#1\endcsname%
    {\the\fnnumber}%
    \insert\footins{\interlinepenalty=\interfootnotelinepenalty
    \leftskip=0pt\rightskip=0pt\parfillskip=0pt plus 1fill
    \spaceskip=2.5pt plus 1.5pt minus 1pt
    \xspaceskip=3pt plus 2pt minus 1pt
    \baselineskip = 0.8\normalbaselineskip \lineskip = \normallineskip
    \pretolerance=2000\tolerance=3000
    \hyphenpenalty=500\exhyphenpenalty=50
    \hangindent 4.1mm\hangafter 1
    \noindent\hbox to 4.1mm{\hss$^{\sixrm\the\fnnumber}$\kern 0.67mm}%
    \vrule width 0pt height 1\baselineskip\egtrm #2}%
    \fi$^{\csname FnN#1\endcsname}\!$}

\title{A Search for the Electric Dipole Moment of the $\tau$-Lepton}
\vspace*{2mm}
\inst {The ARGUS Collaboration}

\authors {%
 H\:Albrecht,
 T\:Hamacher,
 R\:P\:Hofmann,
 T\:Kirchhoff,
 R\:Mankel\fn{Kola}{Now at Institut f\"ur Physik,
                 Humboldt-Universit\"at zu Berlin, Germany.},
 A\:Nau,
 S\:Nowak\fn{IfH}{DESY, IfH Zeuthen, Germany.},
 D\:Re\ss ing,
 H\:Schr\"oder,
 H\:D\:Schulz,
 M\:Walter\fn{IfH}{},
 R\:Wurth
}
\inst {DESY, Hamburg, Germany}

\authors {%
 C\:Hast,
 H\:Kapitza,
 H\:Kolanoski\fn{Kola}{},
 A\:Kosche,
 A\:Lange,
 A\:Lindner,
 M\:Schieber,
 T\:Siegmund,
 H\:Thurn,
 D\:T\"opfer,
 D\:Wegener}
\inst {Institut f\"ur Physik\fn{DO}{%
 Supported by the German
 Bundesministerium f\"ur Forschung und Technologie, under contract
 number 054DO51P.},
 Universit\"at Dortmund,
 Germany}

\authors {%
 C\:Frankl,
 J\:Graf,
 M\:Schmidtler\fn{Sci}{Now at Caltech, Pasadena, USA.},
 M\:Schramm,
 K\:R\:Schubert,\\
 R\:Schwierz,
 B\:Spaan,
R\:Waldi}
\inst{Institut f\"ur Kern- und Teilchenphysik\fn{DR}{%
Supported by the German Bundesministerium f\"ur Forschung
und Technologie, under contract number 056DD11P.},
Technische Universit\"at Dresden, Germany}

\authors {%
 K\:Reim,
 H\:Wegener}
\inst{Physikalisches Institut\fn{ER}{%
 Supported by the
 German Bundesministerium f\"ur Forschung und Technologie, under
 contract number 054ER12P.}, Universit\"at Erlangen-N\"urnberg,
 Germany}

\authors {%
 R\:Eckmann,
 H\:Kuipers,
 O\:Mai,
 R\:Mundt,
 T\:Oest,
 R\:Reiner,
 A\:Rohde,
 W\:Schmidt-Parzefall}
\inst{II. Institut f\"ur Experimentalphysik, Universit\"at Hamburg,
 Germany}

\authors {%
 J\:Stiewe,
 S\:Werner}
\inst{Institut f\"ur Hochenergiephysik\fn{HD}{%
 Supported by the
 German Bundesministerium f\"ur Forschung und Technologie, under
 contract number 055HD21P.}, Universit\"at Heidelberg, Germany}

\authors {%
 K\:Ehret,
 W\:Hofmann,
 A\:H\"upper,
 K\:T\:Kn\"opfle,
 J\:Spengler}
\inst{Max-Planck-Institut f\"ur Kernphysik, Heidelberg, Germany}

\authors {%
 P\:Krieger%
\fn{To}{University of Toronto, Toronto, Ontario, Canada.},
 D\:B\:MacFarlane%
\fn{Mg}{McGill University, Montreal, Quebec, Canada.},
 J\:D\:Prentice\fn{To}{},
 P\:R\:B\:Saull\fn{Mg}{},
 K\:Tzamariudaki\fn{Mg}{},
 R\:G\:Van~de~Water\fn{To}{},
 T.-S\:Yoon\fn{To}{}}
\inst {Institute of Particle
 Physics\fn{CDN}{%
 Supported by the Natural Sciences and Engineering
 Research Council, Canada.}, Canada}

\authors {%
 M\:Schneider,
 S\:Weseler}
\inst{Institut f\"ur Experimentelle Kernphysik\fn{KA}{%
 Supported by the
 German Bundesministerium f\"ur Forschung und Technologie, under
 contract number 055KA11P.}, Universit\"at Karlsruhe, Germany}

\authors {%
 M\:Bra\v cko,
 G\:Kernel,
 P\:Kri\v zan,
 E\:Kri\v zni\v c,
 G\:Medin\fn{lj}{On leave from University of Montenegro, Yugoslavia},
 T\:Podobnik,
 T\:\v Zivko
}
\inst{Institut J. Stefan and Oddelek za fiziko\fn{YU}{Supported
 by the Ministry of Science and Technology of the Republic of
 Slovenia and the Internationales B\"uro KfA,
 J\"ulich.}, Univerza v Ljubljani, Ljubljana, Slovenia}

\authors {%
 V\:Balagura,
 S\:Barsuk,
 I\:Belyaev,
 R\:Chistov,
 M\:Danilov,
 V\:Eiges,
 L\:Gershtein,
 Yu\:Gershtein,
 A\:Golutvin,
 O\:Igonkina,
 I\:Korolko,
 G\:Kostina,
 D\:Litvintsev,
 P\:Pakhlov,
 S\:Semenov,
 A\:Snizhko,
 I\:Tichomirov,
 Yu\:Zaitsev}
\inst{Institute of Theoretical and Experimental Physics\fn{ITEP}{%
 Partially supported by Grant MSB300 from the International
Science Foundation.},
Moscow, Russia}

\vskip 0pt plus 1fill
\eject
}%
\setcounter{page}{1}
\parindent0.0pt
\renewcommand{\arraystretch}{1.2}
\renewcommand{\topfraction}{0.9999}
\renewcommand{\bottomfraction}{0.9999}
\renewcommand{\textfraction}{0.0}
\begin{abstract}
Using the ARGUS detector at the $e^{+}e^{-}$ storage ring DORIS II, we
have searched for the real and imaginary part of the electric dipole formfactor
$d_{\tau}$ of the $\tau$ lepton in the production of 
$\tau$ pairs at $q^2=100 \GeV^2$.
This is the first direct measurement of this $\cal CP$ 
violating formfactor.
We applied the method of optimised observables which takes into account all available information on the observed $\tau$ decay 
products. 
No evidence for $\cal CP$ violation was found, 
and we derive the following results:
${\cal R}e (d_{\tau})=(1.6\pm1.9)\cdot 10^{-16}\ecm$ and 
${\cal I}m (d_{\tau})=(-0.2\pm0.8)\cdot 10^{-16}\ecm$,
where statistical and systematic errors have been combined.
\end{abstract}
\vspace{0.5cm}

The understanding of $\cal CP$ violation and the question if the current description of the electroweak interaction in the mimimal Standard Model (MSM) is sufficient to explain the matter-antimatter asymmetry 
in the universe is one of the burning issues of present particle physics.
It appears that there is insufficient $\cal CP$ violation in the MSM to generate the observed baryon asymmetry \cite{antimatter}, stimulating the search for additional sources 
of $\cal CP$ violation beyond the MSM, e.~g.~leptoquarks, additional 
Higgs bosons or supersymmetric particles.

One way to search for new $\cal CP$ violating interactions is in the pair production of $\tau$ leptons. Contributions from such interactions 
are parametrised model independently 
by the electric dipole formfactor $d_\tau(q^2)$.
Up to now,
upper limits on the weak dipole formfactor have been determined 
at LEP \cite{LEP}, and indirect measurements 
exist of the electric dipole formfactor \cite{lepdip}.


The $\tau$ dipole formfactor is measured in charge dependent momentum correlations in the final states of the reaction $e^+e^-\rightarrow\tau^+\tau^-\rightarrow f^+f^-$. At low energies, 
$\tau$ pairs are produced in the $\cal CP$ even state $^3S_1$ via a virtual photon with quantum numbers $J^{PC}=1^{--}$. $\cal CP$ 
violation would result in the $\cal CP$ odd state $^1P_1$ which leads to other correlations between $\vec{p}(f^+)$ and $\vec{p}(f^-)$ than in the state $^3S_1$.

In the Born approximation, the Lorentz invariant
matrix element for the production of $\tau$ pairs in the reaction 
$e^+e^-\rightarrow \gamma^*\rightarrow\tau^+\tau^-$ 
including $\cal CP$-violating terms is  given by \cite{overmann,graf}
\begin{eqnarray*}
|{\cal{M}_{\rm prod}}|^2 &\propto&
(p_1 q_2) (p_2 q_1)+(p_1q_1)(p_2q_2)+m_{\tau}^2(p_1p_2) \nonumber \\
&+&(\omega_1\omega_2)\left ( (p_1q_2)(p_2q_2)-\frac{1}{2}m_{\tau}^2(p_1p_2)\right ) \\ &-& (\omega_1p_1)(\omega_1p_2)(p_2q_2)-(\omega_1p_1)(\omega_2p_2)(p_1q_2) \nonumber \\
&+& {\cal{R}}e (d_{\tau}) \left (2~ m_{\tau}~ \epsilon_{\mu\nu\rho\sigma}~ (p_1+p_2)^{\mu} \omega_{1}^{\rho} \omega_{2}^{\sigma} \right .\nonumber \\
& & \phantom{\cal{R}e (d_{\tau})}~\left .
\left ( p_{1}^{\nu} \left ( (p_2q_1)-(p_1q_1)\right )-q_1^{\nu}(p_1p_2)\right )\right ) \nonumber \\
&+& {\cal{I}}m (d_{\tau}) \left (\left ( m_{\tau}^2 (\omega_1-\omega_2)_{\mu} 
\left ( -(p_1q_1)p_1^{\mu}+(p_2q_1)p_1^{\mu}+(p_1q_1)p_2^{\mu}-(p_2q_1)p_2^{\mu} \right ) \right . \right .\\
& & \phantom{{\cal{I}}m (d_{\tau})}~+
(q_1q_2)\left ( (p_1q_1)(-(\omega_1p_1)+(\omega_2p_1)+(\omega_1p_2)-(\omega_2p_2)) \right. \nonumber \\
& & \phantom{{\cal{I}}m (d_{\tau})}~\phantom{(q_1q_2)}~+\left .
(p_2q_1)((\omega_1p_1)-(\omega_2p_1)-(\omega_1p_2)+(\omega_2p_2))
\right ) 
\nonumber \\
& & \phantom{{\cal{I}}m (d_{\tau})}~- 
(\omega_2q_1)\left ((p_1q_1)^2+(p_2q_1)^2\right ) + (\omega_1q_2) \left ( ( p_1q_1)^2+(p_2q_1)^2\right ) \nonumber \\
& & \left .\left. \phantom{{\cal{I}}m (d_{\tau})}~+ 
2^{\phantom{2}}(p_1q_1)(p_2q_1)\left((\omega_2q_1)+(\omega_1q_2)\right) \right) \right )\nonumber \\
&+&|d_{\tau}|^2~\left ( -m_{\tau}^2 (p_1p_2)-(p_1q_1)^2+2(p_1q_1)(p_2q_1)-(p_2q_1)^2+(p_1p_2)(q_1q_2) \right ) \nonumber \\
& & \phantom{d_{\tau}^2}~
\left (m_{\tau}^2 +(q_1q_2)-(\omega_2q_1)(\omega_1q_2)+m_{\tau}^2(\omega_1\omega_2)+(q_1q_2)(\omega_1\omega_2) \right )\quad , \nonumber
\end{eqnarray*}

where $d_\tau (s)$ is the electric dipole formfactor of the $\tau$
 lepton, $p_{1,2}$ ($q_{1,2}$) are the 4-momenta of $e^+, e^-$ 
($\tau^+,\tau^-$),  and $\omega_{1,2}$ denote 
the 4-polarisations of $\tau^+$ and $\tau^-$.

Because of the long lifetime of the $\tau$-lepton,  the matrix elements for the production of a $\tau$ pair, $ |{\cal M}_{\rm prod}|^2$,
and that of the subsequent decays, $|{\cal M}_{\rm dec}|^2$, factorize in the Born approximation:
\begin{eqnarray*}
|{\cal M}|^2&=& |{\cal M}_{\rm prod}|^2 \cdot |{\cal M}_{\rm dec}|^2.
\end{eqnarray*}
$ |{\cal M}_{\rm prod}|^2$ shows a linear dependance on ${\cal{R}}e (d_{\tau})$, ${\cal{I}}m (d_{\tau})$, and
$|d_{\tau}|^2$,  whereas  $|{\cal M}_{\rm dec}|^2$ is independent of the electric dipole moment. Therefore,
the differential cross section of the process 
$e^+e^-\rightarrow\tau^+\tau^-\rightarrow f^+f^-$ will exhibit a similar 
linear dependance :
\begin{displaymath}
d\sigma_{\rm tot} = d\sigma_{\rm SM} + 
\Re (d_\tau) d\tilde\sigma_{\cal R}+
\Im (d_\tau) d\tilde\sigma_{\cal I} + |d_\tau|^2 d\tilde\sigma_{\rm Q},
\end{displaymath}
where $d\sigma_{\rm SM}$ corresponds to the differential cross section as expected in the Standard Model. 
Since $|d_\tau|^2 \cdot s\ll1$,
$|d_\tau|^2 d\tilde\sigma_{\rm Q}$ can be neglected for 
$|d_\tau|<5\cdot10^{-16}$ ecm, and the differential cross section
can be rewritten as:

\begin{eqnarray*}
d\sigma_{\rm tot} = d\sigma_{\rm SM}\cdot [ 1+ 
\Re (d_\tau)\omega_{\cal R}+\Im (d_\tau)  \omega_{\cal I}]\quad,&
\omega_{\cal R}=\displaystyle{\frac{d\tilde\sigma_{\cal R}}{d\sigma_{\rm SM}}}\quad ,& 
\omega_{\cal I}=\frac{d\tilde\sigma_{\cal I}}{d\sigma_{\rm SM}}\quad .\\
\end{eqnarray*}

In this form, $\omega_{\cal R}$ and $\omega_{\cal I}$ are the {\sl so-called} optimised observables as suggested in ref.\ \cite{overmann}. 
Both quantities $\omega_{\cal R}$ and $\omega_{\cal I}$ describe $\cal CP$ asymmetries with the properties $\int \omega_{\cal R}d\sigma_{SM} = \int \omega_{\cal I}d\sigma_{SM} = 0$ and $\int \omega_{\cal R}\omega_{\cal I}d\sigma_{SM} = 0$. This leads to   

\begin{displaymath}
\Re(d_{\tau})=\frac{\bra \omega_{\cal R}\ket}{\bra \omega_{\cal R}^2\ket}\quad ,\qquad \Im(d_{\tau})=\frac{\bra \omega_{\cal I}\ket}{\bra \omega_{\cal I}^2\ket}\quad ,
\end{displaymath}

where the means are defined by 
\begin{displaymath}
\bra \omega_j \ket = \frac{\int \omega_j d\sigma_{\rm tot}}{\int d\sigma_{\rm tot}} \quad ,
\end{displaymath}
with $j = {\cal R}, {\cal I}$, and $\bra \omega_j^2 \ket$ accordingly. For the
statistical errors, error propagation gives: 

\begin{eqnarray*}
\delta \Re(d_{\tau}) = \frac{1}{\sqrt{N \bra \omega_{\cal R}\ket}}\sqrt{1-\Re(d_{\tau}^2)\frac{\bra \omega_{\cal R}^4\ket}{\bra \omega_{\cal R}^2\ket}}\quad , &&
\delta \Im(d_{\tau}) = \frac{1}{\sqrt{N \bra \omega_{\cal I}\ket}}\sqrt{1-\Im(d_{\tau}^2)\frac{\bra \omega_{\cal I}^4\ket}{\bra \omega_{\cal I}^2\ket}}\quad . \\
\end{eqnarray*}
Experimentally, the only accessible means are
\begin{displaymath}
\bra \omega_j \ket_{\rm exp} = \frac{\int \omega_j\ \eta\ d\sigma_{\rm tot}}{\int \eta\ d\sigma_{\rm tot}} \quad ,
\end{displaymath} 
$\bra \omega_j^2 \ket_{\rm exp}$ accordingly, and real and imaginary part of $d_\tau$ are determined by $\bra \omega_j \ket_{\rm exp}/\bra \omega_j^2 \ket_{\rm exp}$.
The influence of the detector acceptance $\eta$ has to be determined by Monte Carlo
simulation and will be discussed later in this paper.

The following analysis is limited to the final states $\tau\tau\rightarrow(\rho\nu)(\rho\nu)$, $(\rho\nu)(\mu\nu\nu)$ and $(\rho\nu)(e\nu\nu)$ since these are the most sensitive
ones given the ARGUS environment.
The application of the method of optimised observables requires the integration of each observed event
over all unmeasured quantities, such as $\tau$ direction, photons of the initial state bremsstrahlung which mostly escape undetected, radiated photons in the decay $\tau^\pm\rightarrow l^\pm\nu\bar{\nu}$ that merge in the cluster of the charged particle $l^\pm$, and photons from external bremsstrahlung. 

The integration is performed numerically using Monte Carlo methods. $500$ tries are made for each event using a hit and miss 
approach for the relevant kinematical quantities. 
As in refs.~\cite{moritz} and \cite{schramm}, 
where this method has been used for the determination of Michel parameters, 
there are $n_{\rm hit}$ successful 
tries in which the observables are compatible with 
the kinematics of a $\tau\tau\rightarrow \gamma_1 f_1 f_2 (\gamma_2)$ 
event. The resulting differential cross section 
for each observed event is then the mean of these $n_{\rm hit}$ values: 
\begin{displaymath}
d\sigma(d_{\tau}) = \frac{1}{n_{\rm hit}}\sum_i d\sigma_i(d_{\tau})\quad .
\end{displaymath}

The validity of the method for determining $d_\tau$ has been tested by generating events with the KORALB/TAUOLA Monte Carlo program \cite{koralb}. To describe the effect of an electric dipole formfactor, these events were weighted with the full matrix element \cite{overmann}. Details can be found in ref.~\cite{graf}.


The analysed data sample has been collected between 1983 and 1989 with the ARGUS detector at the $e^+e^-$-storage ring DORIS II. The detector and its trigger requirements are described in ref.\ \cite{argusdet}. The integrated luminosity for this analysis was $291\pb^{-1}$ corresponding to $277 500$ $\tau$ pairs produced. 


The $\rho\rho$ event selection follows closely the one presented in ref.\ \cite{thurn} and starts with requiring two oppositely charged tracks forming a vertex in the interaction region and with $-0.9 < \cos(\vec{p}_1\vec{p}_2) < -0.5$. This reduces the background from more collinear QED events, like Bhabha and $\mu$ pair events, and takes into account that the deacy products of the two $\tau$ are in opposite hemispheres. Each charged track must have a transverse momentum above $60\MeV/c$ and point into the barrel region ($|\cos\theta|<0.7$) to ensure good trigger conditions. 
Neutral pions are reconstructed by their $\gamma\gamma$ decays. Photons are 
required to deposit at least $100\MeV$ in the calorimeter.
The number of photons in the event has to be 2, 3, or 4. Each of these photons 
must have an angle between $10^\circ$ and $90^\circ$ to one of the two tracks. 
The $10^\circ$ cut is against faked photons from charged particle split-offs. 
The photons are then assigned to their closest track.
Only one or two photons are allowed per track.
Depending on the number of detected photons, neutral pions were 
reconstructed in two ways.
If only one photon was found, a minimum energy of $1\GeV$
was required for this photon to be taken as
{\sl single-cluster-$\pi^0$}.
In case of two photons the system had to fulfill $\chi^2<9$ when kinematically constrained to the $\pi^0$ mass.
An additional cut was made on the combination of the charged track and the reconstructed $\pi^0$ to fulfill $m_{\pi^\pm\pi^0}<1.8\GeV/c^2$.
Two-photon and QED reactions are suppressed by the following requirement:
\begin{eqnarray*}
|\sum \vec{p}_T| c /E_{\rm cms} &>& ( 0.45\cdot \sum |\vec{p}|c /E_{\rm cms}-0.5)^2+0.01)\quad ,
\end{eqnarray*}   
where $\vec{p}$ denotes the momentum and $\vec{p}_T$ denotes the transverse momentum
of charged and neutral particles.

\begin{figure}[h]
\centering
\subfigure[]{\epsfig{file=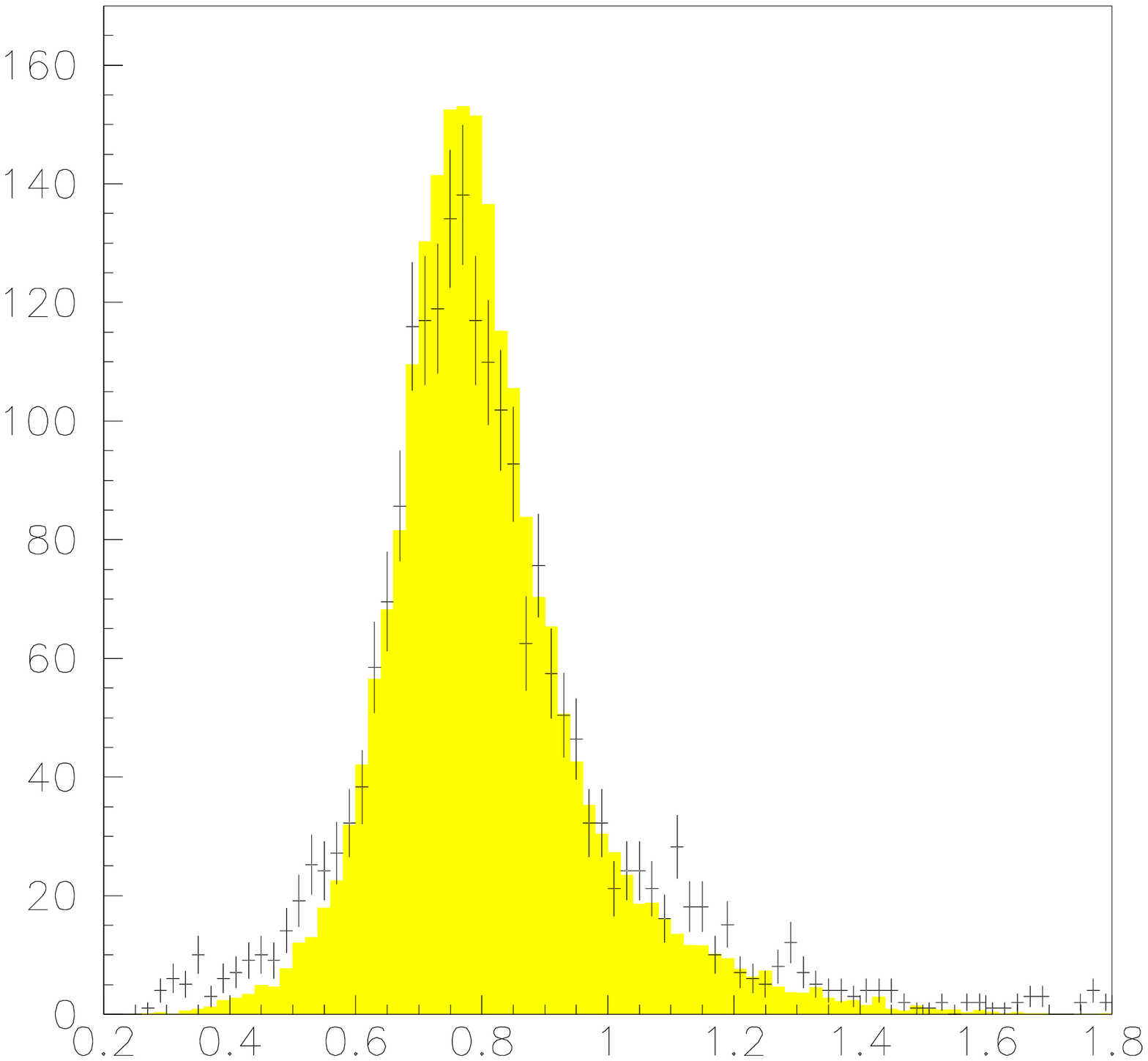,height=6cm,width=6cm,clip=,bbllx=-10,bblly=-10,bburx=560,bbury=535}}
\put(-185,145){\makebox(0,0)[cc]{{\small $\frac{N}{0.02\GeV/c^2}$}}}
\put(-10,5){\makebox(0,0)[cc]{{\small $m_{\pi^\pm\pi^0} [\GeV/c^2]$}}}
\qquad
\subfigure[]{\epsfig{file=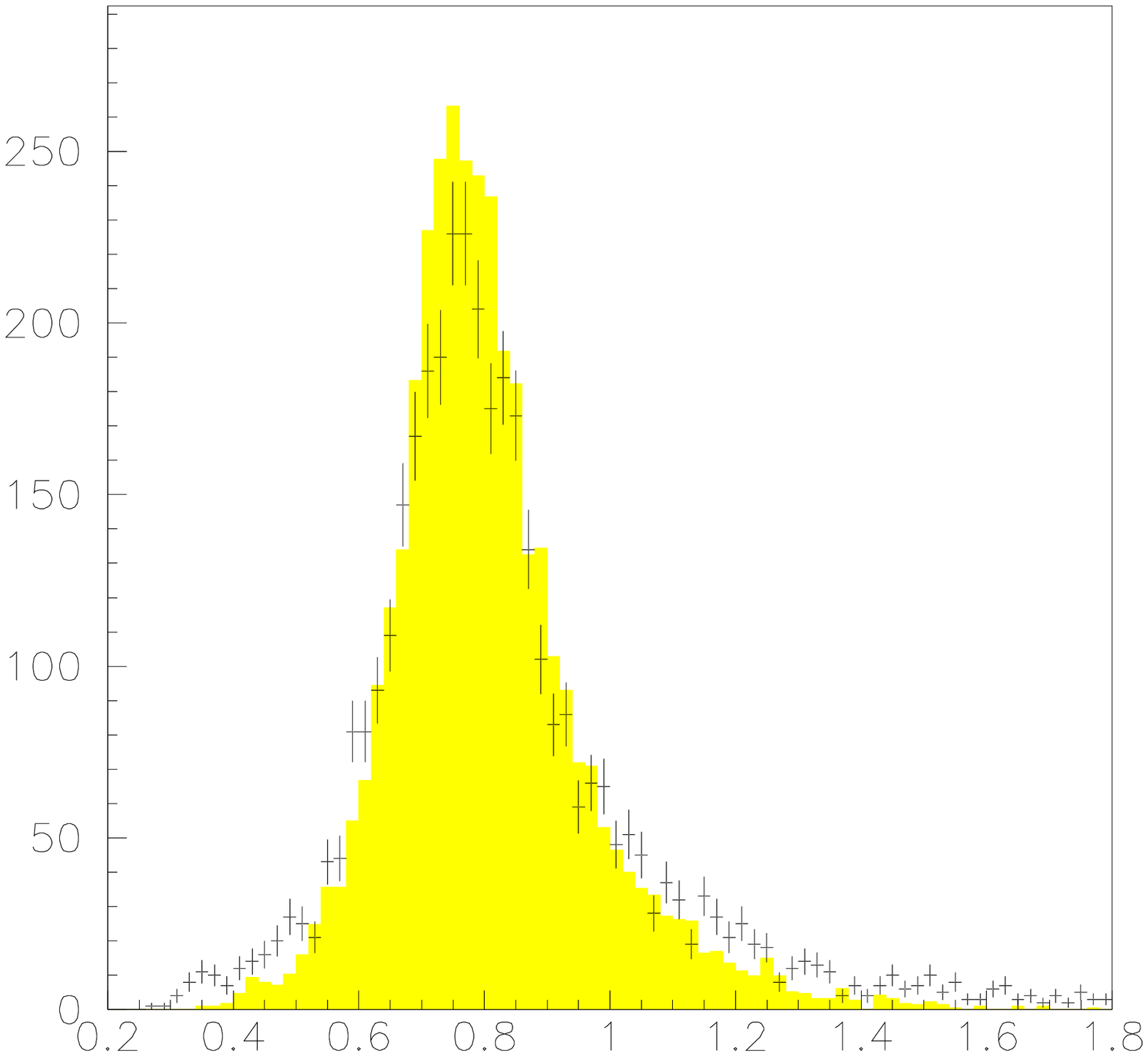,height=6cm,width=6cm,clip=,bbllx=-10,bblly=-10,bburx=560,bbury=535}}
\put(-185,145){\makebox(0,0)[cc]{{\small $\frac{N}{0.02\GeV/c^2}$}}}
\put(-10,5){\makebox(0,0)[cc]{{\small $m_{\pi^\pm\pi^0} [\GeV/c^2]$}}}
\caption{$\pi^\pm\pi^0$ mass spectra for selected (a) $(\rho^\pm\bar{\nu})(\rho^\mp\nu)$ and (b) $(\mu^\pm\nu\bar{\nu})(\rho^\mp\nu)$ events. The points with error bars represent the data.  The hatched histograms show the expectation from the K{\small ORAL}B/T{\small AUOLA} Monte Carlo.\label{rhomass}}
\end{figure}

The $\rho l$ selection was performed like in ref.\ \cite{schramm}. We only present here the differences to the $\rho\rho$ selection. The opening angle of the charged tracks must fulfill $-0.965 < \cos(\vec{p}_1\vec{p}_2) < 0.2$. This cut is much weaker because the background of hadronic events is much less in this decay channel. We also use two different $|\sum \vec{p}_T|$ requirements depending on the $\pi^0$ reconstruction:
\begin{eqnarray*}
{\rm single~cluster:}\quad|\sum \vec{p}_T|c/E_{\rm cms}&>&5\cdot ( M_{\rm miss}^2/E^2_{\rm cms}-0.45)^4 \quad ,\\
{\rm two~cluster:}\quad |\sum \vec{p}_T|c/E_{\rm cms}&>&3\cdot ( M_{\rm miss}^2/E^2_{\rm cms}-0.4)^4\phantom{5} \quad , 
\end{eqnarray*}
where $ M^2_{\rm miss} = (E_{\rm cms}-\sum E_{\rm i})^2-(\sum \vec{p}_{\rm i})^2$.
To ensure a good lepton identification, a momentum above $0.5\GeV/c$ for electrons and above $1.1\GeV/c$ for muons was required. A possible lepton candidate had to fulfill the following particle identification cuts: for muons 
$L_{\mu} > 0.8,  L_{e} < 0.2$, and for electrons 
$L_{e} > 0.8,  L_{\mu} < 0.2$, where the likelihood ratios $L_{e,\mu}$ are defined in ref.~\cite{argusdet}.
In Fig.~\ref{rhomass}, the resulting $\pi\pi$ mass distributions for accepted 
events are compared to Monte Carlo calculations for pure signal events. 
The differences between those spectra are mainly due to $\tau$ background from
the decays $\tau\tau\rightarrow
(\rho\nu)(\pi\pi^0\pi^0\nu)$ and
$\tau\tau\rightarrow(l\nu\nu)(\pi\pi^0\pi^0\nu)$.

\begin{table}[h]
\begin{center}
\begin{tabular}{|c|c|}\hline
\multicolumn{2}{|c|}{$(\rho\nu)(\rho\nu)$} \\ \hline
$(\pi^\pm\pi^0\pi^0\nu)(\rho^\mp\nu)$ & $13\%$\\
$(\mu^\pm\nu)(\rho^\mp\nu)$ & $2\%$\\
$(K^{*\pm}\nu)(\rho^\mp\nu)$ & $2\%$\\ 
$q\bar{q}$ events             & $2\%$ \\
$\gamma\gamma$ events         & $6\%$ \\ \hline
\multicolumn{2}{|c|}{$(\rho\nu)(\mu\nu\nu)$}\\  \hline
$(\pi^\pm\pi^0\pi^0\nu)(\mu^\mp\nu\nu)$ & $4\%$\\
$(K^{*\mp}\nu)(\mu^\pm\nu\nu)$ & $3\%$ \\ \hline
\multicolumn{2}{|c|}{$(\rho\nu)(e\nu\nu)$}\\ \hline
$(\pi^\pm\pi^0\pi^0\nu)(e^\mp\nu\nu)$ & $4\%$\\
$(K^{*\mp}\nu)(e^\pm\nu\nu)$ & $4\%$ \\ \hline
\end{tabular}
\caption{Monte Carlo results for the background estimation. \label{background}}
\end{center}
\end{table}

\begin{figure}[h]
\centering
\subfigure[]{\epsfig{file=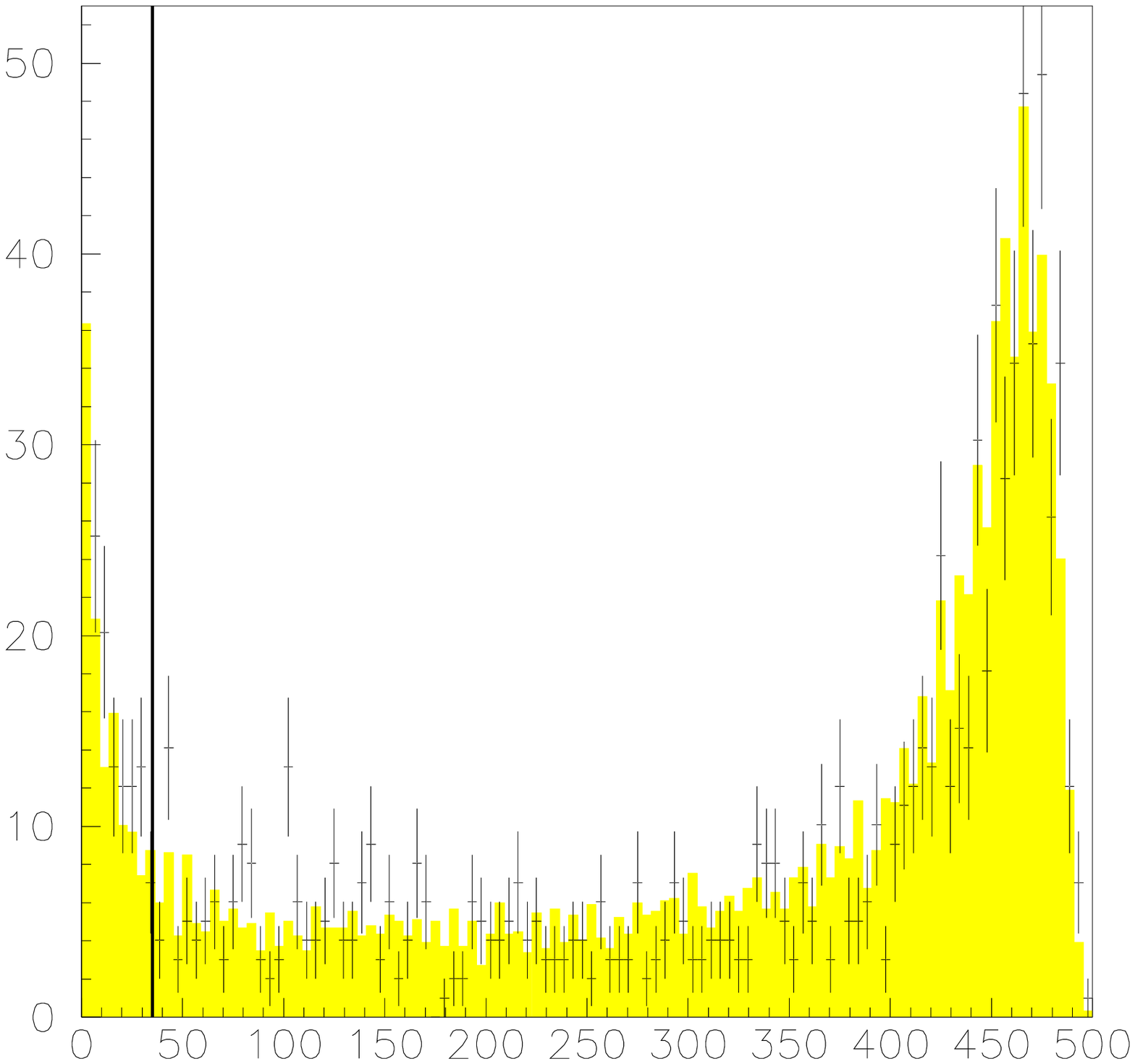,height=4.5cm,width=5.5cm,clip=,bbllx=-10,bblly=-10,bburx=560,bbury=535}}
\put(-155,115){\makebox(0,0)[cc]{{\small $N$}}}
\put(-10,5){\makebox(0,0)[cc]{{\small $n_{\rm hits}$}}}
%
\qquad
\subfigure[]{\epsfig{file=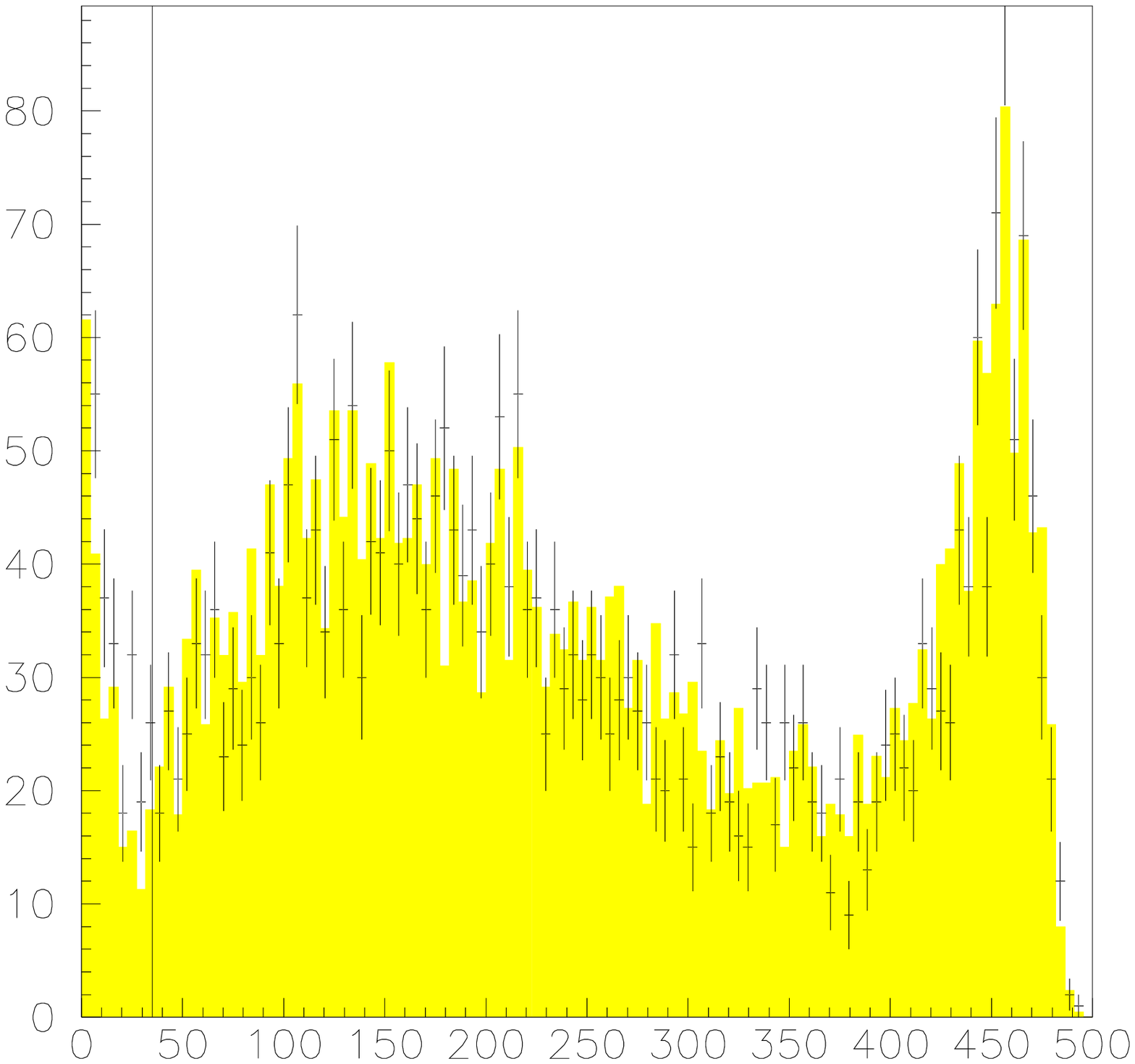,height=4.5cm,width=5.5cm,clip=,bbllx=-10,bblly=-10,bburx=560,bbury=535}}
\put(-155,115){\makebox(0,0)[cc]{{\small $N$}}}
\put(-10,5){\makebox(0,0)[cc]{{\small $n_{\rm hits}$}}}
\caption{Distribution of $n_{\rm hit}$ for (a) $(\rho^\pm\bar{\nu})(\rho^\mp\nu)$ and (b) $(\mu^\pm\nu\bar{\nu})(\rho^\mp\nu)$ where $n_{\rm hit}$ is the number of possible kinematic configurations in 500 tries. The points with error bars represent the selected data. The hatched histograms show the expectations from K{\small ORAL}B/T{\small AUOLA}. A cut
of $n_{\rm hit}>35$ is applied, lower $n_{\rm hit}$ values originate mostly 
from events with hard initial state radiation. \label{rhoistat}}
\end{figure}

The distribution of $n_{\rm hit}$ is shown in Fig.~\ref{rhoistat}. The good agreement between $N(n_{\rm hit})$ for selected data and accepted Monte Carlo events supports the validity of the assumptions used. The high value of $\bra n_{\rm hit}\ket$ shows the effectiveness of the numerical integration.
A cut of $n_{\rm hit}>35$ is applied. 
After the selection we get $1074$ $(\rho\nu)(\rho\nu)$, $2052$ $(\rho\nu)(\mu\nu\nu)$, and $3675$ $(\rho\nu)(e\nu\nu)$ candidates at a mean center of mass energy of $\sqrt{s}=10.38\ \GeV$, defining the $q^2$ scale for the measurement of
$d_\tau$.

\begin{figure}[h]
\centering
\subfigure[]{\epsfig{file=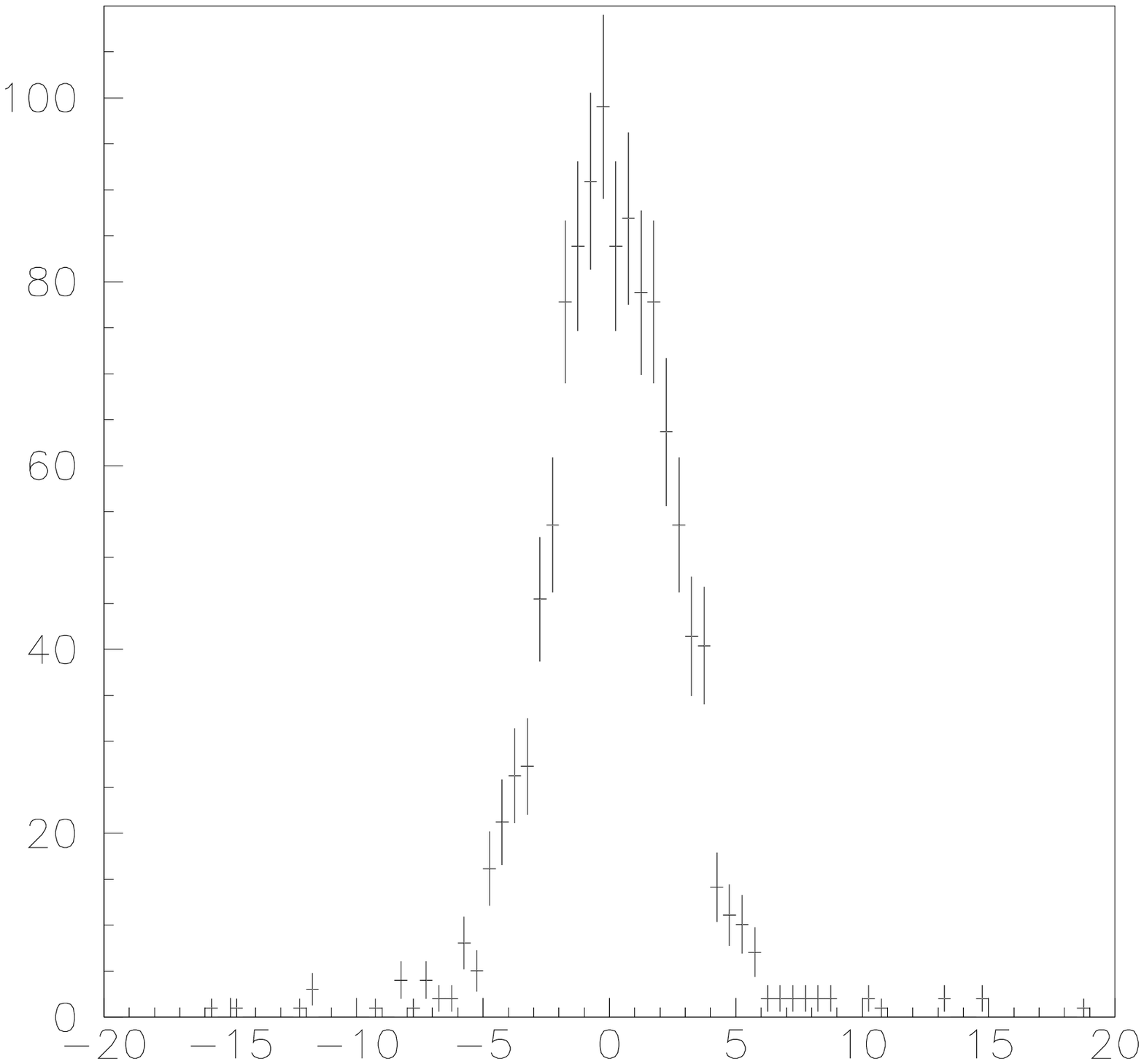,height=6.0cm,width=6.0cm,clip=,bbllx=-10,bblly=-10,bburx=560,bbury=560}} 
\subfigure[]{\epsfig{file=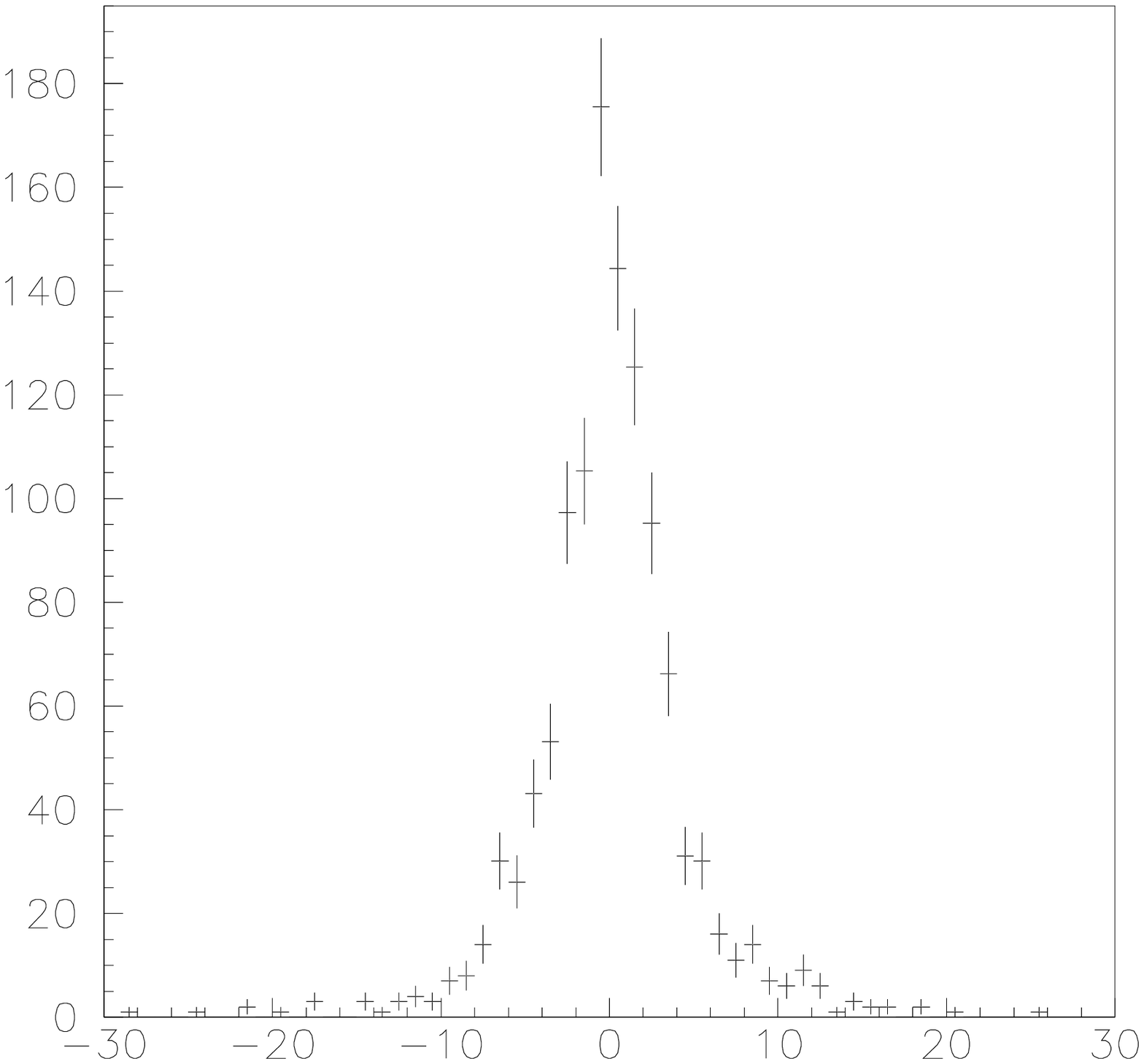,height=6.0cm,width=6.0cm,clip=,bbllx=-10,bblly=-10,bburx=560,bbury=560}}
\caption{Optimised observables in the decay $\tau\tau\rightarrow(\rho\nu)(\rho\nu)$\label{rhorhoreaima}}
\unitlength=1mm
\begin{picture}(0,0)
\put(-60,84){\makebox(0,0)[cc]{{\small $\frac{N}{0.5\GV}$}}}
\put(-15,28){\makebox(0,0)[cc]{{\small ${\omega}_{\cal R}[\GV]$}}}
\put(2,84){\makebox(0,0)[cc]{{\small $\frac{N}{\GV}$}}}
\put(45,28){\makebox(0,0)[cc]{{\small ${\omega}_{\cal I}[\GV]$}}}
\end{picture}
\end{figure}

\begin{figure}[h]
\centering
\subfigure[]{\epsfig{file=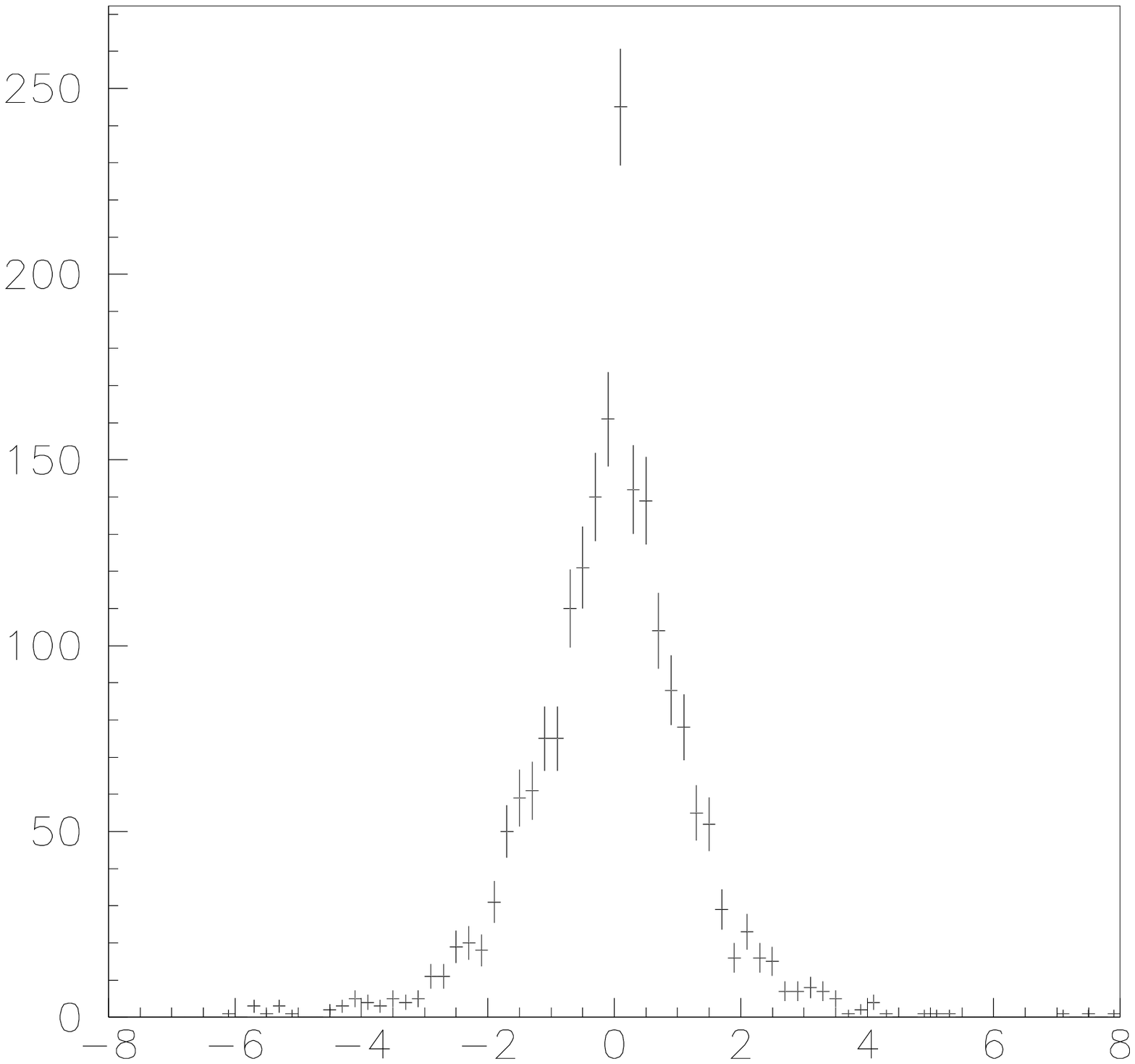,height=6.0cm,width=6.0cm,clip=,bbllx=-10,bblly=-10,bburx=560,bbury=560}} 
\subfigure[]{\epsfig{file=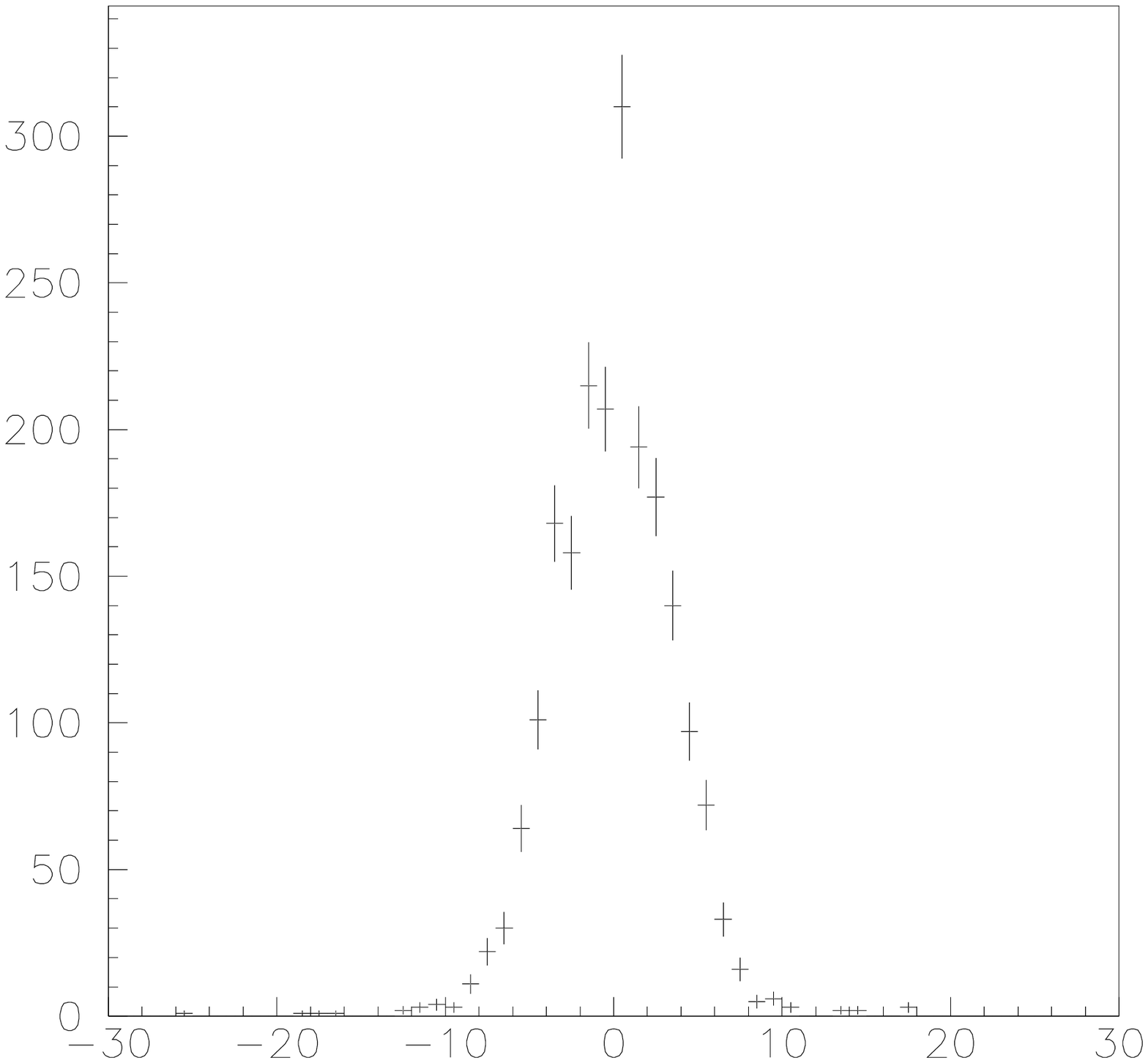,height=6.0cm,width=6.0cm,clip=,bbllx=-10,bblly=-10,bburx=560,bbury=560}}
\caption{Optimised observables in the decay $\tau\tau\rightarrow(\rho\nu)(\mu\nu\nu)$\label{rhomureaima}}
\unitlength=1mm
\begin{picture}(0,0)
\put(-62,84){\makebox(0,0)[cc]{{\small $\frac{N}{0.2\GV}$}}}
\put(-15,28){\makebox(0,0)[cc]{{\small ${\omega}_{\cal R}[\GV]$}}}
\put(0,84){\makebox(0,0)[cc]{{\small $\frac{N}{\GV}$}}}
\put(45,28){\makebox(0,0)[cc]{{\small ${\omega}_{\cal I}[\GV]$}}}
\end{picture}
\end{figure}

\begin{figure}[h]
\centering
\subfigure[]{\epsfig{file=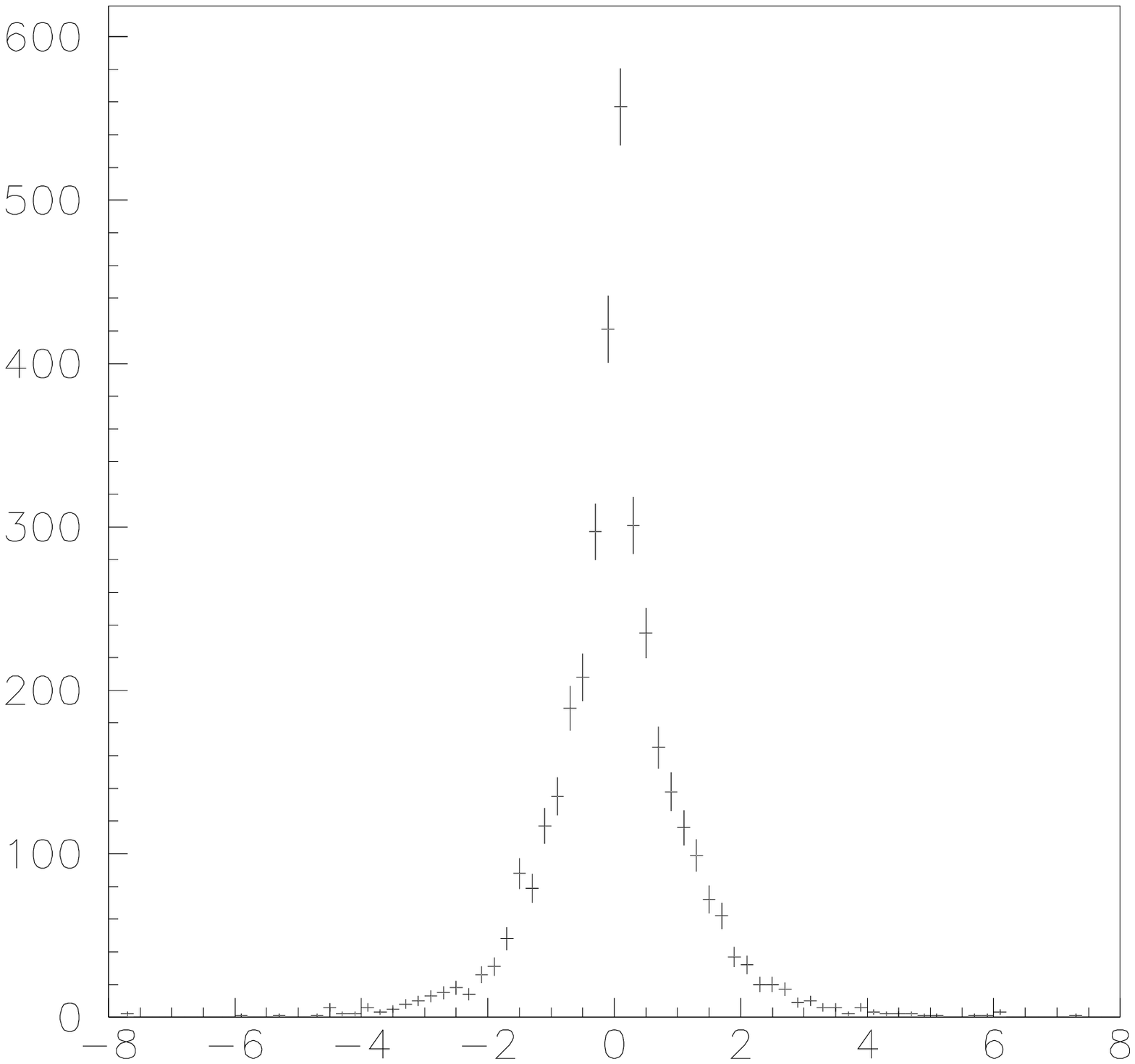,height=6.0cm,width=6.0cm,clip=,bbllx=-10,bblly=-10,bburx=560,bbury=560}} 
\subfigure[]{\epsfig{file=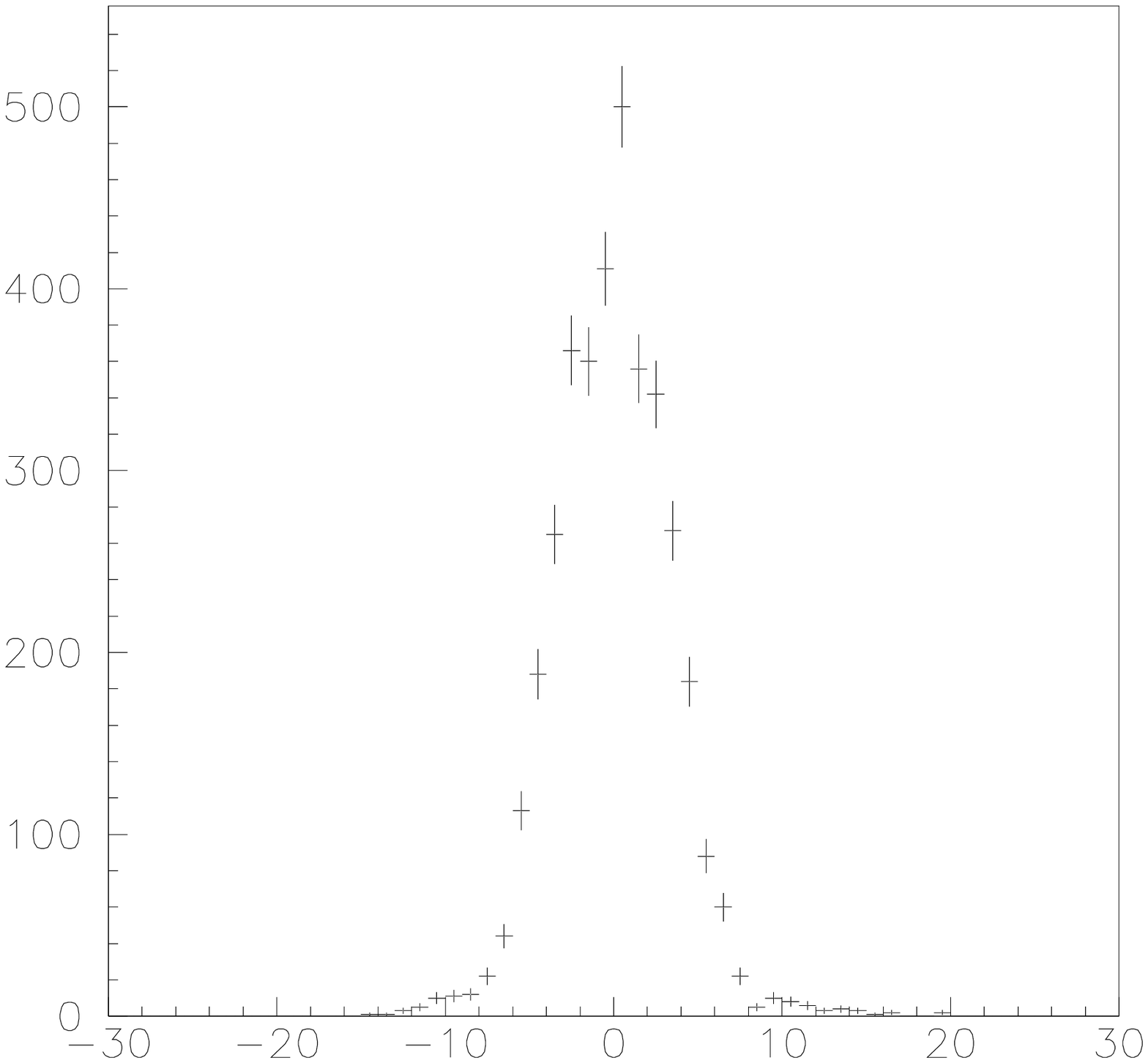,height=6.0cm,width=6.0cm,clip=,bbllx=-10,bblly=-10,bburx=560,bbury=560}}
\caption{Optimised observables in the decay $\tau\tau\rightarrow(\rho\nu)(e\nu\nu)$\label{rhoereaima}}
\unitlength=1mm
\begin{picture}(0,0)
\put(-62,84){\makebox(0,0)[cc]{{\small $\frac{N}{0.2\GV}$}}}
\put(-15,28){\makebox(0,0)[cc]{{\small ${\omega}_{\cal R}[\GV]$}}}
\put(0,84){\makebox(0,0)[cc]{{\small $\frac{N}{\GV}$}}}
\put(45,28){\makebox(0,0)[cc]{{\small ${\omega}_{\cal I}[\GV]$}}}
\end{picture}
\end{figure}

The selected events lead to distributions of the optimised observables as 
shown in Figs.~\ref{rhorhoreaima}, \ref{rhomureaima}, and \ref{rhoereaima}. 
The first and second moment of these distributions are used to determine the electric dipole formfactor. 
All six distributions are centered around zero 
indicating $d_\tau$ close to zero.
The widths of the three $\omega_{\cal I}$ distributions are very similar. 
Note that the 
width of the $\omega_{\cal R}$ distribution is largest for the 
$\rho\rho$ channel. This reflects the higher information on $\Re(d_{\tau})$
per event.

The results have still to be corrected for systematic uncertainties, 
like acceptance corrections, particle identification, and background. 
The influence of background channels was studied using Monte Carlo 
calculations. For the $(\rho\nu)(e\nu\nu)$ and $(\rho\nu)(\mu\nu\nu)$ 
channels the background of hadronic,
 gamma-gamma and QED events (Bhabha, $\mu$-pair) was found to be negligible. 
The only background source is from non-signal $\tau$ events. 
In the $(\rho\nu)(\rho\nu)$ channel, the background 
from gamma-gamma and hadronic events has to be considered. 
The resulting background contributions are summarised 
in Table \ref{background}.

The influence of the $\tau$ background on $d_\tau$ was studied with event samples generated with $d_{\tau}=0$. This is possible because of the small amount of background and the loss of spin information in most channels, e.~g.~\ in the decay $\tau^\pm\rightarrow \pi^\pm\pi^0\pi^0\nu$ where some photons or a $\pi^0$ were not detected.   
The Monte Carlo results were used to correct the obtained values for $\Re(d_{\tau})$ and $\Im(d_{\tau})$. 
Results after correction are given in Table 3 with their statistical errors.

\begin{table}[h]
\begin{center}
\begin{tabular}{|c|ccc|ccc|}\hline
systematic error source & \multicolumn{3}{c|}{real part $[10^{-16}\ecm]$} & \multicolumn{3}{c|}{imaginary part $[10^{-16}\ecm]$} \\ 
        & $\delta_{\Re}^{\rho\rho}$ & $\delta_{\Re}^{\rho\mu}$ & $\delta_{\Re}^{\rho e}$ & $\delta_{\Im}^{\rho\rho}$ & $\delta_{\Im}^{\rho\mu}$ & $\delta_{\Im}^{\rho e}$ \\ \hline
acceptance correction & $0.48$  & $1.00$  & $1.00$  & $0.35$  & $0.38$  & $0.36$ \\
charged asymmetrie    & $-$     & $-$     & $-$     & $-$     & $0.24$  & $0.06$\\
lepton identifikation & $-$     & $0.022$ & $0.010$ & $-$     & $0.007$ & $0.006$ \\ 
photon identification & $0.005$ & $0.002$ & $0.002$ & $0.003$ & $0.002$ & $0.001$ \\
background variation& $0.06$ & $0.13$ & $0.001$ & $0.02$ & $0.02$ & $0.003$ \\ \hline
total systematic error & $0.48$ & $1.01$ & $1.00$ & $0.35$ & $0.45$ & $0.37$ \\ \hline
\end{tabular}
\caption{Contributions to the systematic error.\label{syserror}}
\end{center}
\end{table}

Detector acceptance effects have been studied in a Monte Carlo simulation by generating large event samples with $\Re (d_\tau)$ and $\Im (d_\tau)$ equal to $1\cdot 10^{-16}$ and $5\cdot 10^{-16}\ecm$. After reconstruction including all acceptance cuts, these values were obtained again within uncertainties listed as systematic 
error in the first line of Table 3. Strong influences on the $d_\tau$ measurement may occur if the detector acceptance is different for particles of opposite charge. Therefore, this effect was extensively studied with simulated data where holes in momentum, polar and azimuthal angle were generated. No charge dependent effects were found.
In addition,
systematic errors in the determination of the momentum of charged
particles showed no influence on the $d_\tau$ determination.
A detailed study was necessary because charge dependent 
systematic momentum shifts of up to 
$\pm$20 MeV/$c$ for momenta above
4 GeV/$c$ have been observed for some data taking periods. 


The systematic error estimates are given in Table \ref{syserror}, their sums also in Table \ref{ergebnis}.
No evidence for $\cal CP$ violation was found.
One of the three final states ($\rho\nu\rho\nu$) was also investigated with a maximum likelihood fit method.
The result of this fit was identical to the one with optimised observables, the
statistical errors were equal within $1 \%$.
\begin{table}[h]
\begin{center}
\begin{tabular}{|l|ccc|} \hline
channel & \multicolumn{2}{|c|}{dipole formfactor} & value $[10^{-16}\ecm]$ \\ \hline \hline
$\rho\rho$ & $\Re(d_{\tau})$ &$=$& $\phantom{-}3.3\pm2.5\pm0.5$ \\
           & $\Im(d_{\tau})$ &$=$& $\phantom{-}0.5\pm1.4\pm0.4$ \\ \hline
$\rho\mu$  & $\Re(d_{\tau})$ &$=$& $-4.0\pm5.3\pm1.0$ \\
           & $\Im(d_{\tau})$ &$=$& $-1.2\pm1.8\pm0.5$ \\ \hline
$\rho  e$  & $\Re(d_{\tau})$ &$=$& $\phantom{-}0.2\pm3.1\pm1.0$ \\
           & $\Im(d_{\tau})$ &$=$& $-0.2\pm1.1\pm0.4$ \\ \hline

\end{tabular}
\caption{Final results with statistical and systematic errors for all three selected channels.\label{ergebnis}}
\end{center}
\end{table}

\newpage
With statistical and systematic errors added in quadrature, the combination of the three measurements leads to: 

\begin{eqnarray*}
\Re(d_{\tau}) &=& (\phantom{-}1.6\pm 1.9)\cdot 10^{-16}\ecm \quad ,\\
\Im(d_{\tau}) &=& (-0.2\pm0.8)\cdot 10^{-16}\ecm \quad ,
\end{eqnarray*}

which translates into the following upper limits:

\begin{eqnarray*}
|\Re(d_{\tau})| &<& 4.6 \cdot 10^{-16}\ecm \qquad (95\% {\rm CL})\quad , \\
|\Im(d_{\tau})| &<& 1.8 \cdot 10^{-16}\ecm \qquad (95\% {\rm CL})\quad . \\
\end{eqnarray*}

The result is in agreement with current LEP results \cite{lepdip}, which used the assumption of a non-existing anomalous magnetic dipole moment to calculate an upper limit on the electric dipole moment.

\section*{Acknowledgements}
We would like to thank O. Nachtmann and P. Overmann for their stimulation and
helpful discussions.
We also thank U. Djuanda, E. Konrad, E. Michel, and W. Reinsch 
for their competent technical help in running the experiment and processing
the data, as well as Dr. H. Nesemann, B. Sarau, and the DORIS group for
the operation of the storage ring. The visiting groups wish to thank
the DESY directorate for the support and kind hospitality extended to them.


\begin{thebibliography}{99}
\bibitem{antimatter} E.~W.~Kolb, M.~S.~Turner, 
The Early Universe, Addison Wesley (1994).
\bibitem{LEP} OPAL Collaborarion, K. Ackerstaff et al.,  
Z. Phys. {\bf C74} (1997) 403;\\
ALEPH Collaboration, D. Buskulic et al., 
Phys.~Lett.~{\bf B}$\bf 346$ (1995) 371; \\
L3 Collaboration, M.~Acciarri et al., 
Phys.~Lett.~{\bf B}$\bf 426$ (1998) 207.
\bibitem{lepdip} L3 Collaboration, M.~Acciarri et al., 
Phys.~Lett.~{\bf B}$\bf 434$ (1998) 169; \\
OPAL Collaboration, K.~Ackerstaff et al., 
Phys.~Lett.~{\bf B}$\bf 431$ (1998) 188. 
\bibitem{overmann} W. Bernreuther, O. Nachtmann, P. Overmann, 
Phys.Rev.{\bf D}$\bf 48$ (1993) 78.
\bibitem{graf} J. Graf,
``Suche nach CP-verletzenden Korrelationen in der Tau-Paarproduktion
bei einer Schwer-punktsenergie von $\sqrt{s}=10~\rm GeV"$, 
Dr. rer. nat. Thesis, Technische Universit\"at Dresden, TUD-IKTP/99-05 (1999).
\bibitem{moritz} ARGUS Collaboration, 
H. Albrecht et al., Phys. Lett. {\bf B349} (1995) 576.
\bibitem{schramm} ARGUS Collaboration, 
H. Albrecht et al., Phys. Lett. {\bf B431} (1998) 179.
\bibitem{koralb}S.~Jadach, Z.~ W\c as, R.~Decker, 
Comput.~Phys.~Commun.~$\bf 36$ (1985) 191; \\
S.~Jadach, Z.~ W\c as, R.~Decker, 
Comput.~Phys.~Commun.~$\bf 64$ (1991) 275;\\
S.~Jadach, Z.~ W\c as, R.~Decker, J.~H.~K\"uhn, 
Comput.~Phys.~Commun.~$\bf 76$ (1993) 361.
\bibitem{argusdet} 
ARGUS Collaboration, H.Albrecht et al.,
Nucl. Instr. Meth. {\bf A275} (1989) 1.
\bibitem{thurn} ARGUS Collaboration, H. Albrecht et al.,
Phys. Lett. {\bf B337} (1994) 383.

\end{thebibliography}
\end{document}